\documentclass[conference]{IEEEtran}
\IEEEoverridecommandlockouts
\usepackage{cite}
\usepackage{dsfont}
\usepackage[cmex10]{amsmath}

\usepackage{multirow}
\usepackage{booktabs,colortbl}

\usepackage[cmex10]{amsmath}
\usepackage{siunitx}
\usepackage{cite}
\usepackage{psfrag}
\usepackage[utf8]{inputenc}
\usepackage[T1]{fontenc}
\usepackage{amsmath,amsfonts,amsbsy,amssymb}
\usepackage{mathabx}
\usepackage{mathrsfs}
\usepackage[nolist]{acronym}
\usepackage{tabularx}
\usepackage{amssymb}
\usepackage{amsmath}
\usepackage{graphicx}
\usepackage{cite}
\usepackage{multirow}
\usepackage{wasysym}
\usepackage{multirow}
\usepackage{float}
\usepackage{color}
\usepackage{algorithmic}
\usepackage{graphicx}
\usepackage{textcomp}
\usepackage[font=small,skip=0pt]{caption}

\newtheorem{proposition}{Proposition}

\newtheorem{remark}{Remark}
\def\BibTeX{{\rm B\kern-.05em{\sc i\kern-.025em b}\kern-.08em
		T\kern-.1667em\lower.7ex\hbox{E}\kern-.125emX}}
\begin{document}

\title{Energy Efficiency of an Unlicensed Wireless Network in the Presence of Retransmissions}
\author{\IEEEauthorblockN{Iran Ramezanipour\IEEEauthorrefmark{1}, Hirley Alves\IEEEauthorrefmark{1}, Pedro. H. J. Nardelli\IEEEauthorrefmark{2} and Ari Pouttu\IEEEauthorrefmark{1}}
	%
	\IEEEauthorblockA{\IEEEauthorrefmark{1}Centre for Wireless Communications (CWC), University of Oulu, Finland\\
	}
	\IEEEauthorblockA{
		\IEEEauthorrefmark{2}Lappeenranta University of Technology, Lappeenranta, Finland\\
	}
	Firstname.lastname@oulu.fi, Firstname.lastname@lut.fi 
}

	%
	
	
%

\maketitle

\begin{abstract}
This paper analysis the energy efficiency of an unlicensed wireless network in which retransmission is possible if the  transmitted message is decoded in outage.  A wireless sensor network is considered in which the sensor nodes are unlicensed users of a wireless network which transmit its data in the uplink channel used by the licensed users. Poisson point process is used to model the distributions of the nodes and the interference caused by the licensed users for the sensor nodes. After finding the optimal throughput in the presence of retransmissions, we focus on analyzing the total power consumption and energy efficiency of the network and how retransmissions, network density and outage threshold affects the energy efficiency of the network.

\end{abstract}


\section{Introduction}

Wireless sensor networks (WSNs) have gained a lot of popularity during the past few decades and are being implemented in different applications such as military and medical sectors as a mean for monitoring, processing and disseminating data\cite{pantazis2009energy}. The ease of implementation and being cost and energy efficient are among the reasons that have made WSNs popular. Wireless sensor nodes are small size devices that can create dense networks that are randomly positioned and deployed which makes them a suitable choice for inaccessible locations or disaster relief operations as well. WSNs have different properties and applications compared to the traditional wireless ad hoc networks, hence, protocols and algorithms that are being used in those networks are not valid for WSNs anymore  \cite{akyildiz2002survey}. This opens up a wide field of research regarding WSNs \cite{al2015internet}.

Having an energy efficient network is always challenging when dealing with wireless networks. WSNs are also no exception to this, specially since they usually have access to a limited power source both in terms of the available energy ($<0.5$ Ah, $1.2$ V)  and size \cite{p2, vardhan2000wireless} and in many cases such as aforementioned hardly accessible locations, it is not possible to renew the power sources for sensor networks which are usually batteries, hence, the battery life in such networks play a crucial role in the sensors lifetime which makes the energy consumption of the network elements a very important factor that needs to be considered when dealing with WSNs \cite{akyildiz2002wireless,de2011energy}. 

Although a sensor network consumes energy in all its three areas of responsibilities which were mentioned earlier, it is the data disseminating, which includes both transmitting and receiving data, that consumes the most energy in a WSN. Sensor networks are usually used in short range communications with low data rates and short packet size which makes the RF communications a suitable choice for them \cite{rabaey2000picoradio}. However, designing an energy efficient WSN is always one of the challenges engineers face since the radio technologies are not suitable for being used in all kinds of applications\cite{shih2001physical}. Thus, in this paper, we analyze the energy efficiency of a WSN as part of an unlicensed network which allows for retransmission in case of an outage event happening.

Energy efficiency (EE) studies have become very popular during the past years and researchers have been studying EE in different types of applications. In \cite{hasan2011green}, authors study the reduction of the energy consumption of the whole network while in \cite{wang2010energy,sadek2009energy}, the energy consumption of two non-cooperative and cooperative networks with considering different network densities have been studied. In \cite{wang2010energy}, optimizing the packet size is used as a mean for maximizing the energy efficiency of the two mentioned networks. In \cite{sadek2009energy}, energy efficiency of a cooperative network is studied constrained by an outage threshold. Moreover, EE is investigated in \cite{de2011energy} by setting an end-to-end throughput constraint on the network while in \cite{alves2014outage}, by studying the throughput and outage of a full-duplex and an incremental cooperative half-duplex networks respectively.

While the following studies are important and valuable, they do not consider EE of a sensor network, based on the optimal throughput of the system, constrained by an outage threshold. In this paper, we expand our previous works in \cite{nardelli2016maximizing,tome2016joint} to cover a more generalized model rather than focusing on only the smart grids application. We follow the same model described in \cite{nardelli2012optimal,nardelli2014throughput} where there is possibility for retransmissions of a message in case of an outage happening in the network and it is shown that having a limited number of retransmissions can enhance the spatial throughput and transmission capacity of ad hoc networks. We use the same network model for investigating the EE of the network by first optimizing the link throughput in the system subjected to a minimum outage requirement where an outage event happens if the transmitted message is not decoded correctly or is never received by the receiver. Note that the number of retransmissions is limited and if the message is not received after a certain number of retransmissions, it is dropped.

The rest of the paper is divided as follows.
Section \ref{sec:model} introduces the system model, while Section \ref{sec:opt} details the proposed throughput optimization and energy efficiency analysis.
Section \ref{sec:res} presents the numerical results and Section \ref{sec:concl} concludes this paper.

\section{System model}
\label{sec:model}
\begin{figure}[!t]%
	\centering
	\includegraphics [width=	\columnwidth]{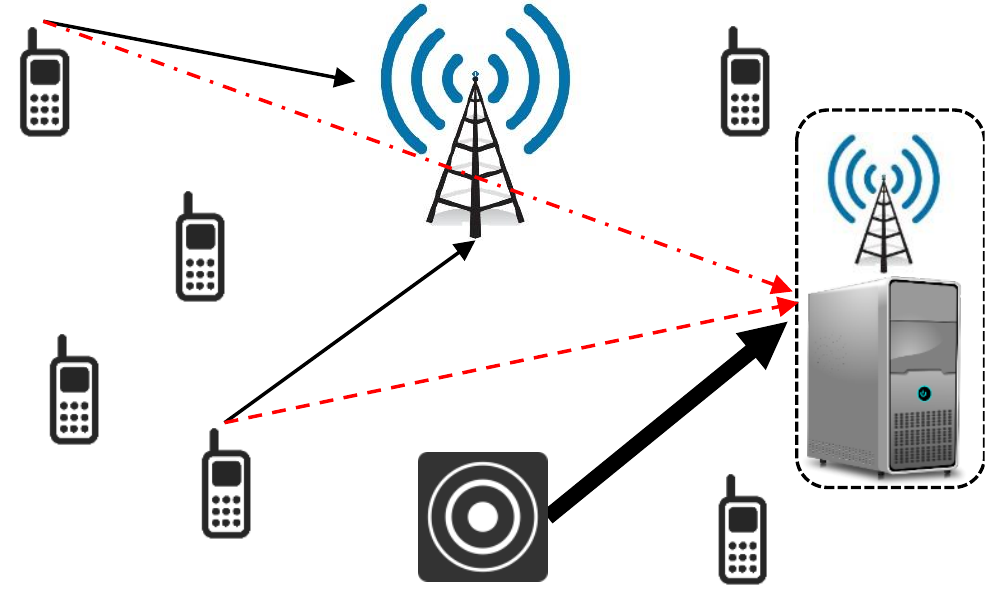}
	%
	\caption{An illustration of the proposed scenario, where licensed and unlicensed users share the up-link channel. The reference sensor (unlicensed transmitter) is depicted by the sensor, the controller (unlicensed receiver) by the CPU and its antenna, the handsets are the mobile licensed users (interferers to the controller) and the big antenna is the cellular base-station. As the sensors uses directional antennas with limited transmit power (bold arrow), its interference towards the base-station can be ignored. The thin black arrows represent the licensed users' desired signal, while the red ones represent their interference towards the controller.}
	\label{fig-scheme}
	%
\end{figure}

%
Considering the network model introduced in \cite{nardelli2016maximizing,tome2016joint}, the same model which is also shown in Fig. \ref{fig-scheme}, is used here for the implementation of the communication network in which the sensors transmit their data to their corresponding aggregator/controller where the following assumptions hold.

\begin{itemize}
\item \textbf{Assumption 1}: A communication network consists of both licensed and unlicensed networks where the users of both of the networks share the frequency bands allocated to the uplink channel.
\item \textbf{Assumption 2}: Licensed link is the connection link between static cellular base-stations and mobile users while the sensor nodes with fixed positions consist of unlicensed users which communicate with their corresponding controller through the uplink channel.
\item \textbf{Assumption 3}: The amount of power used by the unlicensed users (sensors) for their transmission is limited. This limitation can be enforced by the licensed network or can also be related to the sensors' own capabilities.
\item \textbf{Assumption 4}: In this model, it is assumed that there are no packet collisions between sensors associated with the same aggregator/controller due to the fact that the size of the transmitted messages are assumed to be small and multiple access solutions are effective for the size of the unlicensed network.
\end{itemize}

By considering the above assumptions, we are able to simplify the model to some extend. This would make the analysis easier.
The use of directional antennas in unlicensed communication links is justified based on Assumption 2 which states that the unlicensed nodes are static. This would result in not having any orientation errors \cite{wildman2014joint}. Having a limited transmit power based on Assumption 3 means that there is also a limit to the maximum range that the signal sent by the sensors can reach. Hence, the radiation pattern created by this transmission can be modeled as a line segment with the starting point being the sensors and the end point being limited by the imposed power constraint.

The interference in this model based on Assumptions 1, 2 and 4 can have different sources. (i) from unlicensed users (sensors) to cellular base-stations, (ii) from sensors to controller that are not their corresponding controller and (iii) from licensed users (mobile users) to to aggregators/controllers. The first two interference sources can be avoided if when implementing the network, the position of either the licensed or unlicensed nodes are specified. Even if the positions are randomly implemented, it is highly unlikely that there would be a base station or controller in the same line as the sensors' transmitted signal.

This leaves out only one source of interference in this model which would be (iii) from licensed users to aggregators/controllers. In order to be able to analyze the effect of interference on the performance of the network, we need to model the uncertainty of the interferers positions. We use Poisson point process $\Phi$ to model the interfering nodes in this network which are distributed over an infinite two-dimensional plane with network density $\lambda$. Details of using stochastic geometry and Poisson point process in modeling the wireless networks can be found in \cite{haenggi2012stochastic}.

Different metrics such as distance-dependent path-loss and fast fading is considered when modeling the wireless channel. Consider $r_i$ to be the distance between the \textit{i}th interferer and the reference receiver which is located arbitrary at the origin. Based on the Slivynak theorem, a receiver can have an arbitrary fixed position at the center of the Euclidean distance. This would make the estimation of the other elements of the network surrounding the receiver easier \cite{haenggi2012stochastic}. In this model, $g_i$ is considered to be the channel gain. The reference receiver received power then would be $W g_ir_i^{-\alpha}$ where $W$ is the transmit power and $\alpha> 2$ is the path-loss exponent. This will result in the following signal to interference ratio (SIR$_0$).

\begin{equation} 
\label{eq_SIR}
\textrm{SIR}_0 = \dfrac{W_\mathrm{s} g_{0}  r_0^{-\alpha}}{W_\mathrm{p} \; \underset{i \in \Phi}{\displaystyle \sum}  g_{i}  r_i^{-\alpha}}.
\end{equation}

In this equation, $W_\mathrm{p}$ and $W_\mathrm{s}$ denote the licensed users and unlicensed users transmit power respectively.
It should be noted that although the noise is neglected here, even the presence of the noise would not make a qualitative difference as stated also in \cite{weber2010overview}.

Considering that point-to-point Gaussian codes and interference-as-noise decoding rules \cite{nardelli2015throughput,baccelli2011interference} are used in the reference link, it means that obtaining the desired spectral efficiency of $\log_2(1+\beta)$ in bits/s/Hz depends on the fact that the SIR is greater than a given threshold or not $\beta$ (i.e $\textrm{SIR}>\beta$), Hence, the probability of an outage event happening,  $P_\textrm{out}$, can be explained as the probability of $\textrm{SIR}\leq\beta$. If a transmitted message is decoded in outage, it is retransmitted with the maximum of $m$ attempts, meaning that the message is dropped if it is still not successfully decoded by the receiver after $1+m$ transmissions. Thus, the probability of a successful transmission is calculated as $P_\textrm{suc} = 1 - P_\textrm{out}^{1+m}$.

Since the licensed users (interferers) are not static, their position is constantly changing in each transmission. SIR$_0$ in this model can be statistically evaluated by considering different realizations of Poisson point process $\Phi$. In order to compute $P_\textrm{out} = \textup{Pr}\left[\textup{SIR}_0 \leq \beta\right]$ for each transmission attempt by considering quasi-static channel gains (squared envelopes) $g$ which are independent and identically distributed exponential random variables (Rayleigh fading) with mean $1$, the following equation is used \cite{nardelli2012optimal}.

\begin{equation}\label{eq:6}
P_\textrm{out}= 1 - e^{- k \lambda  \beta^{2/\alpha}},
\end{equation}
where $k=\pi r_0^2 \Gamma{\left(1 - \frac{2}{\alpha} \right)} \Gamma{\left(1 + \frac{2}{\alpha} \right)}$.

The throughput of the reference link $T$ is then calculated as \cite{nardelli2012optimal}: 
\begin{equation}\label{eq:5}
T=\frac{\log(1+\beta)}{1+\bar{m}}\left(1-P_\textrm{out}^{1+m}\right),
\end{equation}
where $m$ is the maximum number of retransmission attempts. It should be noted that in order to find $m$, we use an approximation of \cite[§17]{nardellimult2009i} which is also explained in \cite{Iran}, in order to calculate the average number of transmissions needed to successfully transmit a message ($1+\bar{m}$).

\section{Throughput optimization And Energy Efficiency}
\label{sec:opt}
Energy efficiency of a wireless network can be seen as a criteria that captures the trade-off between the total power consumption (PC) and the throughput of the network. Hence, we first start by defining the optimal throughput of the network which is obtained by the following optimization problem: 

\begin{equation}
\begin{aligned}\label{eq:7}
& \underset{(\beta,m)}{\text{max}}
& & \frac{\log(1+\beta)}{1+\bar{m}}\times\left(1-P_\textrm{out}^{1+m}\right)  \\ 
& \text{subject to}
& & P_\textrm{out}^{1+m} \leq \epsilon 
\end{aligned}.
\end{equation}

In this problem, the throughput is constrained to a maximum acceptable error rate $\epsilon$, which shows how often a message is dropped after reaching the maximum number of allowed retransmissions.  Here, the SIR threshold $\beta>0$ and the number of allowed retransmissions $m \in \mathbb{N}$ are the design variables.

%
%
%

\begin{proposition}
The throughput $T = f (\beta, m)$ in \eqref{eq:5} is a function of the variables $m>0$ and $\beta>0$. The function $f$ is then concave with respect to $\beta$ if $\frac{\partial^2 T}{\partial \beta^2} < 0$. 
$\beta^\ast$ and $m^\ast$ represent the value of $\beta$ and maximum number of retransmissions that maximizes the link throughput respectively :
\end{proposition}

%
\begin{equation}\label{eq:8}
\beta^\ast=\left(- \frac{1}{k\lambda} \log{\left (1 - \epsilon^{\frac{1}{m + 1}} \right )}\right)^{\frac{\alpha}{2}}.
\end{equation}

\begin{align}\label{eq:90}
m^\ast=  \max\limits_{m \in \mathbb{N}}  & \;\; \log\left (- \frac{1}{k \lambda} \log\left (1 -\epsilon^{\frac{1}{m + 1}}  \right ) \right ) + \nonumber \\ 
&\alpha  \left(- \frac{1}{k\lambda}\right)^{\frac{\alpha}{2}} \; \frac{ \left(\log\left (1 - \epsilon^{\frac{1}{m + 1}} \right ) \right )^{\frac{\alpha}{2} - 1}}{2 - \frac{2}{k\lambda}  \left (\log\left (1 - \epsilon^{\frac{1}{m + 1}} \right ) \right )^{\frac{\alpha}{2}}} .
\end{align}

%

\begin{IEEEproof}
	As $m$ and $\beta$ are strictly positive variables and function $T$ is twice differentiable in terms of $\beta$, then $T$ is concave if and only if $\frac{\partial^2 T}{\partial \beta^2} < 0$. Eq. \eqref{eq:8} is then attained by solving the derivative equation $\partial T /\partial \beta = 0$, whose solution is $\beta^\ast$. From (\ref{eq:8}), we find $T$ as a function of $m$ considering  $\beta^*$.
	The optimal throughput $T^*$ in terms of both $m$ and $\beta$ is then given by the value of $m$ that maximizes the throughput, which is given in \eqref{eq:90}. Moreover, from (\ref{eq:8}), we find $T$ as a function of $m$ considering  $\beta^*$.
	The optimal throughput $T^*$ in terms of both $m$ and $\beta$ is then given by the value of $m$ that maximizes the throughput, which is given in \eqref{eq:90}.
\end{IEEEproof}
%
%

%
%

%
%

%

\begin{remark} The maximum number of retransmissions $m^*$ is a natural number that is usually small, which makes the evaluation of \eqref{eq:10} computationally simple.

\end{remark}

By having the optimal throughput, we can now analyze the energy efficiency of the network. As it was mentioned earlier, the EE depends on the total power consumption and throughput of the network where the total power consumption of the network in its turn, includes the distance dependent transmission power in addition to the total energy consumed by the RF components and bit rate \cite{de2011energy,alves2014outage}. Considering the above parameters, the total power consumption of our single hop model is derived as:

\begin{equation}
\textup{PC}=\sum_{1}^{m+1} \frac{PC_{PA}+PC_{T_x}+PC_{R_x}}{\textup{log} (1+\beta^*)},
\end{equation}

\noindent where $PC_{PA}$ denotes the power amplifier power consumption in a one-hop transmission which also depends on a parameter called the drain efficiency of the amplifier. We denote the drain efficiency by $\zeta$ which would result in $PC_{PA}=\frac{\beta^*}{\zeta}$ in this model. Moreover, $\textup{log} (1+\beta^*)$ represents the bit rate (bits/s) of the system while $PC_{T_x}$ and $PC_{R_x}$ are constants which depends on the current technologies and are equal to the energy consumed during the transmission and reception operations by the internal circuitry respectively. By having the above parameters, the EE is expressed as

\begin{equation}\label{eq:10}
\textup{EE}=\frac{T^*}{\textup{PC}},
\end{equation}

\noindent where $T^*$ denotes the optimal throughput previously calculated.

\section{Numerical results}
\label{sec:res}

In this part, the numerical results of our analysis is presented. It should be noted that the following parameters were considered for obtaining these results. The distance between the sensors and the receiver $r_0=1$ meter and path-loss exponent $\alpha=4$. Also, based on the parameter setting presented in \cite{de2011energy}, $PC_{T_x}=97.9$ mW, $PC_{R_x}=112.2$ mW and $\zeta=0.35$.

\begin{figure}[!t]
	\includegraphics[width=1.05\columnwidth]{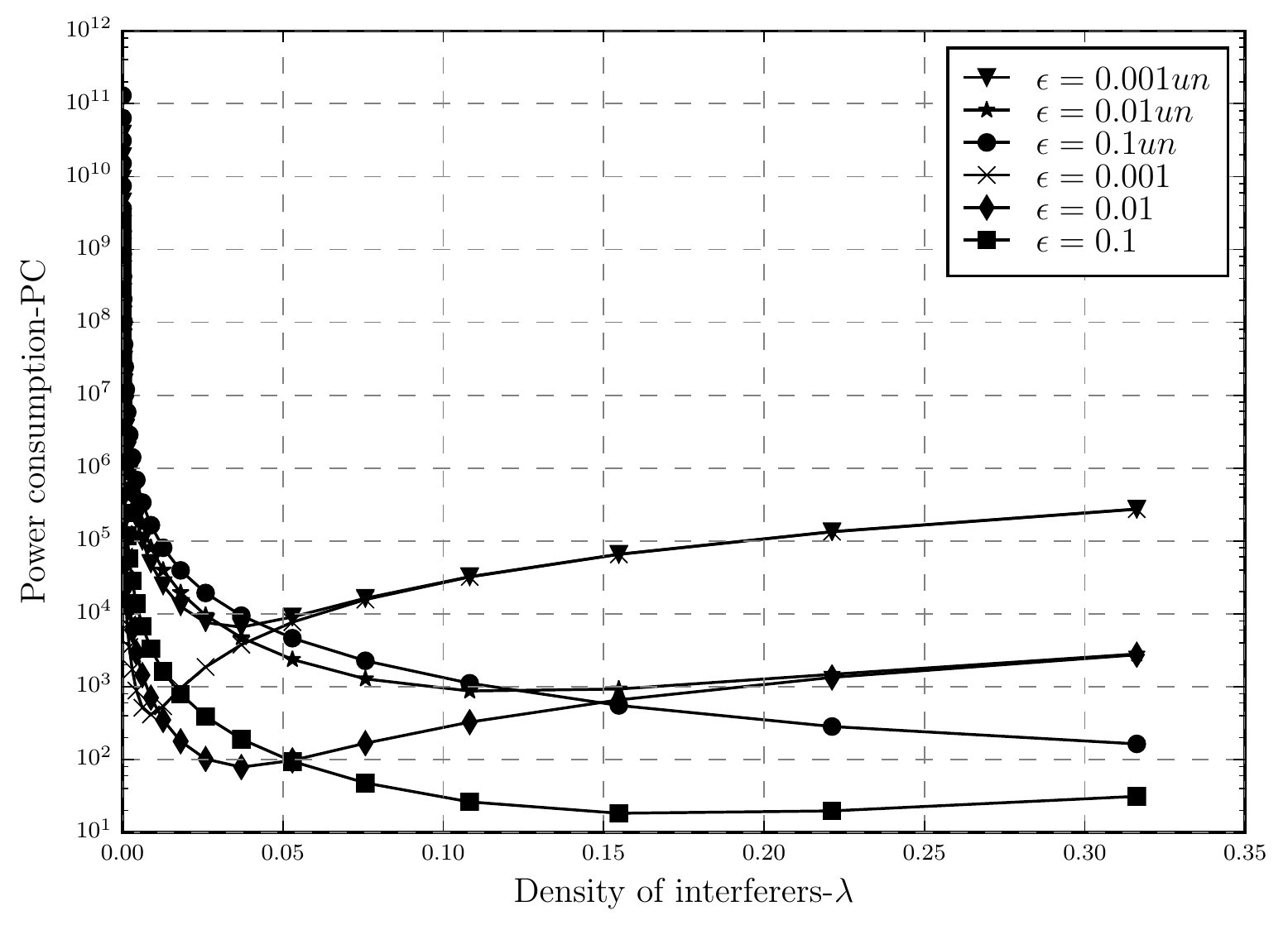}
	\caption{Power Consumption PC versus the density of interferers $\lambda$ for $\alpha=4$, $r_0=1$ for both unlimited and limited ($m=5$) number of retransmissions.}
	\label{fig:pc}
\end{figure}

Fig. \ref{fig:pc} shows how the power consumption changes with the density of interferers (measured in node/$m^2$ ) and outage constraints of the network for both limited and unlimited number of retransmissions being allowed in the network. We can see that in both cases, the outage constraint has a big impact on the total power consumption of the network. As the outage requirement gets stricter, meaning that a lower level of outage is allowed in the network (higher reliability), the total power consumption of the network also increases, showing that more power is used by the network in order to have a successfully decoded transmission.

Moreover, we can see that increasing the density of interferers also affects the total power consumption. While $\epsilon=0.001$, for very low density of interferers ($\lambda\leq 0.07$), having limited number of retransmissions results in having lower total power consumption. However, as the density of interferers increases and expectedly the total power consumption also increases, limited and unlimited retransmissions consume almost the same amount of energy. As the outage requirement gets looser, the range of $\lambda$ for which the limited transmission consumes less energy also increases. For instance, while ($\lambda\leq 0.22$), having limited $m$ means having lower energy consumption when $\epsilon=0.01$. When the outage requirement of the system is very loose, $\epsilon=0.1$, for all of the considered $\lambda$ range in our analysis, having limited $m$ would consume less energy, since when the network density is lower, the interference level of the network is also lower. This means that even when $m$ is limited, the system can achieve its expected outage constraint without having to consume a lot of energy, hence, having limited $m$ consumes less energy, but as $\lambda$ increases, the level of interference also increases which would mean that when having limited $m$, the system needs to use more energy in order to reach the required $\epsilon$, thus, almost the same amount of energy as having unlimited $m$ would be used by the network.

\begin{figure}[!t]
	\includegraphics[width=1.05\columnwidth]{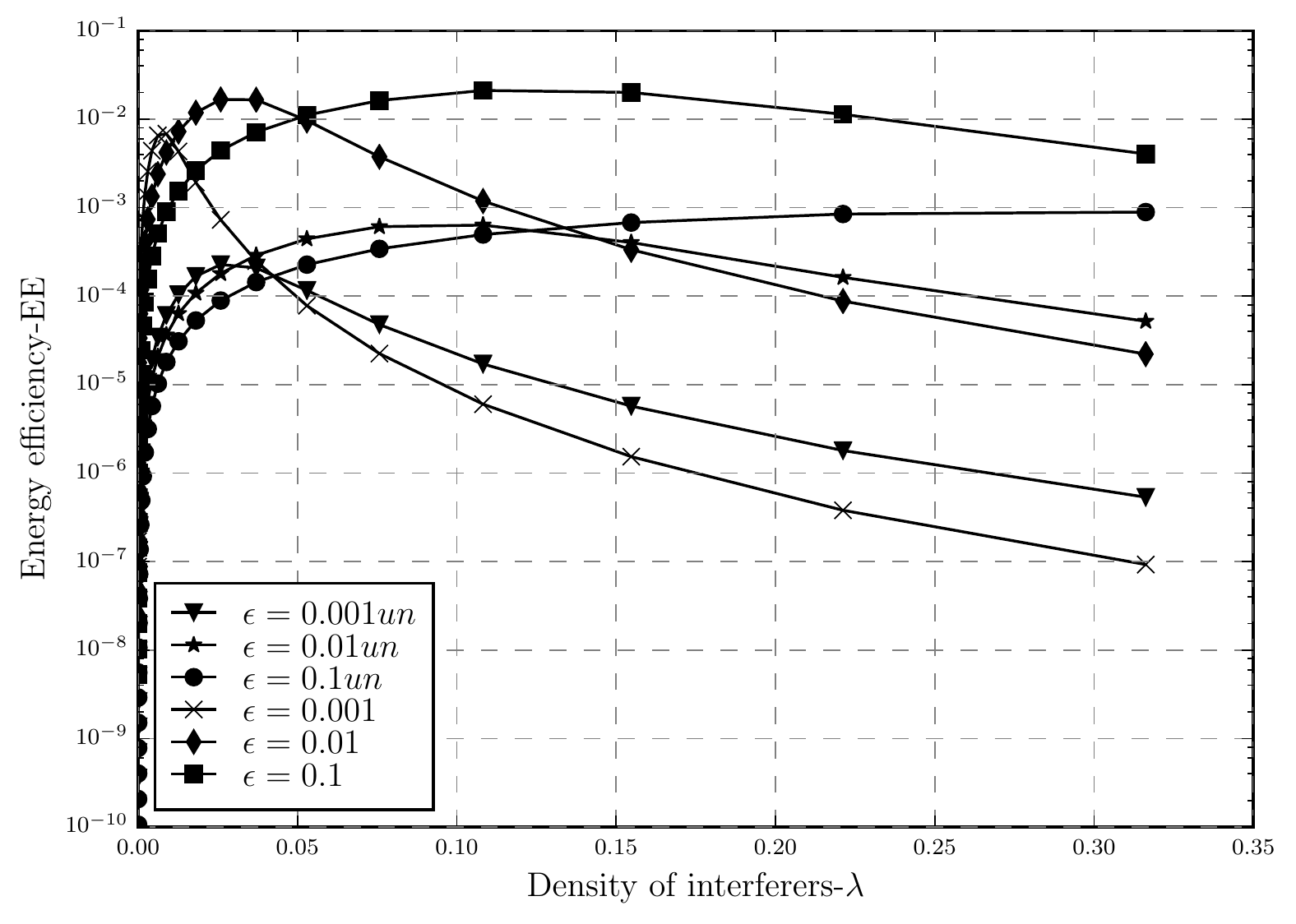}
	\caption{Energy efficiency EE versus the density of interferers $\lambda$ for $\alpha=4$, $r_0=1$ for both unlimited and limited ($m=5$) number of retransmissions.}
	\label{fig:ee}\vspace{-3mm}
\end{figure}

Fig. \ref{fig:ee} illustrates the behavior of the energy efficiency of the network with respect to $\lambda$ and different outage requirements. As the optimal throughput reduces dramatically by $\lambda$, we can see that the same thing is happening in the case of EE. As the outage requirement gets more stringent, the system needs more retransmission in order to reach the optimal throughput. This means that if $m$ is limited, the system can not always reach the optimal throughput as $\lambda$ increases. Since as shown in \eqref{eq:10}, EE has a direct relationship with the optimal throughput, the throughput decrease will effect EE also, that is why we can see that for most of the $\lambda$ range,  having a limited number of retransmissions has also reduced the energy efficiency of the network compared to when an unlimited number of retransmissions is allowed. Although energy efficiency also depends on the power consumption which was shown increases with $\lambda$, this change is not as high and as effective as the decrease is the throughput, hence, the EE eventually ends up decreasing. 

It is also shown in Fig. \ref{fig:ee} that like PC, EE also has a different behavior when $\lambda$ is very low. For those cases, having limited $m$ results in having higher EE. As it was explained earlier, for low $\lambda$, the network would consumes less energy while $m$ is limited, resulting in higher energy efficiency in the network. However, as $\lambda$ increase, limited $m$ uses as much energy as unlimited $m$ in order to reach the required constraints, on the other hand, the throughput for the unlimited case decreases also since the network can not reach the optimal throughput anymore. All these would eventually result in the network having lower EE when the number of retransmissions is limited as $\lambda$ and interference level increase. 

\begin{figure}[!t]
	\includegraphics[width=1.05\columnwidth]{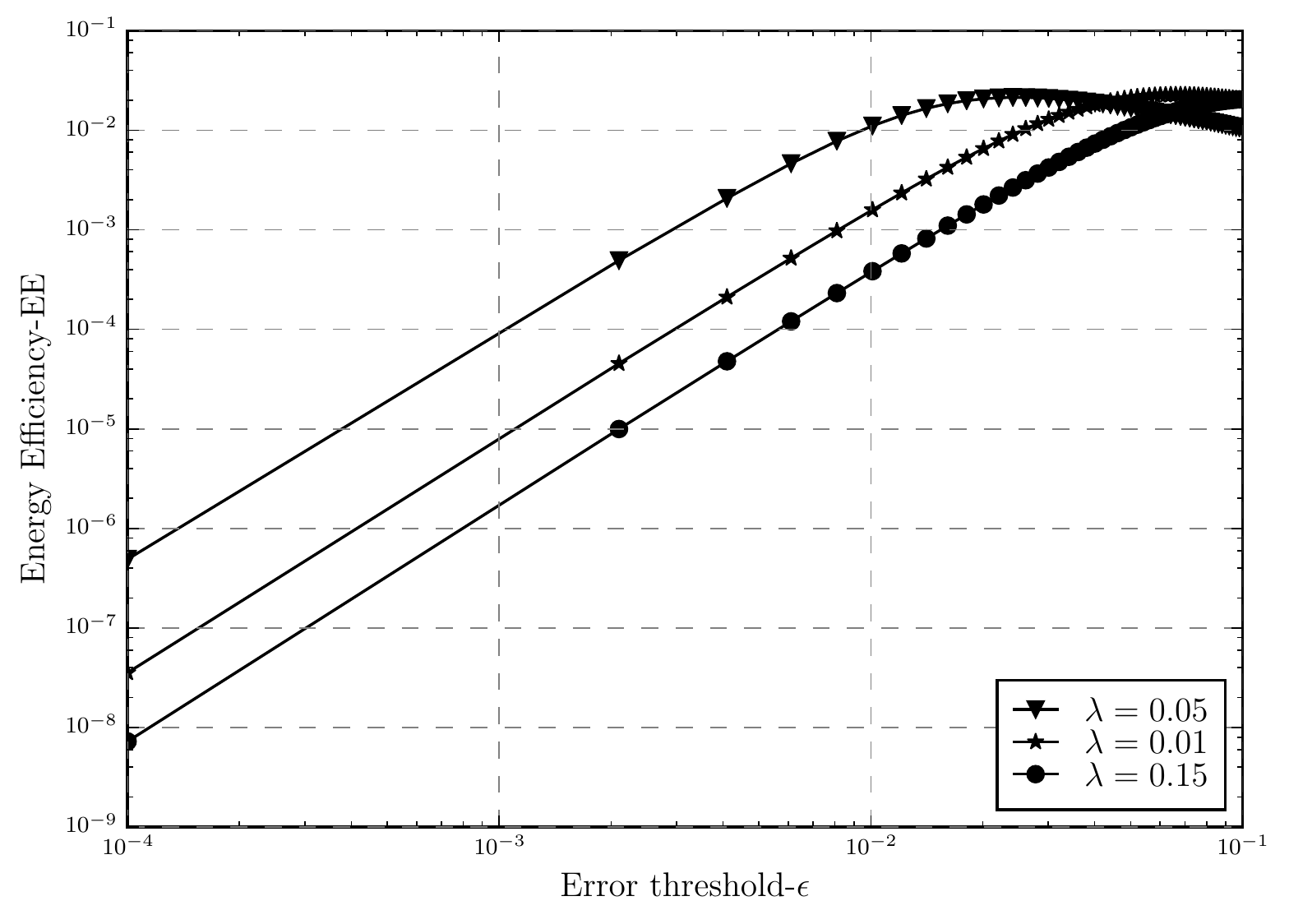}
	\caption{Energy efficiency EE versus the error threshold $\epsilon$ for $\alpha=4$, $r_0=1$ for different densities of interferers $\lambda$ while the number of retransmissions is limited ($m=5$).}
	\label{fig:eree}\vspace{-3mm}
\end{figure}

It should be noted that the sharp fall and rise in Figs. \ref{fig:pc} and \ref{fig:ee} respectively are the results of having very low $\lambda$ which means having a very low interference. Moreover, Fig.\ref{fig:eree} illustrates the EE behavior as a function of $\epsilon$ for different $\lambda$s which further proves our point showing that looser outage requirements and lower network densities results in a higher level of energy efficiency in the network. 
%
%

%

\section{Conclusion}
\label{sec:concl}

In this paper, we analyzed the energy efficiency of a wireless network where the licensed and unlicensed users share the uplink channel. However, the unlicensed users do not cause interference on the licensed users transmissions. In this model, retransmission is also allowed if a message is decoded in outage. Our results showed the effect of retransmission and outage constraint on the power consumption and energy efficiency of the network considering different network densities. It was shown that having stricter outage requirement in the network also means having higher power consumption during transmissions. Depending on $\epsilon$ and $\lambda$, having limited retransmissions means lower power consumption or at most as much power consumption compared to having unlimited $m$. We also showed that as $\lambda$ increases, the energy efficiency of the network decreases due to the decrease in the optimal throughput. Having higher outage requirement also results in needing more $m$ in order to reach the $T^*$, therefore, having limited $m$ means having lower $EE$ for most of the $\lambda$ range. We plan to continue and improve the work done in this paper by jointly optimizing the energy efficiency and throughput constrained by a minimum outage requirement.

\section*{Acknowledgments}

This work is partially supported by Aka Project SAFE (Grant n.303532) and Strategic Research Council/Aka BCDC Energy (Grant n.$292854$).

\bibliographystyle{IEEEtran}
\bibliography{IEEEabrv,refsURC}

\begin{thebibliography}{10}
\providecommand{\url}[1]{#1}
\csname url@samestyle\endcsname
\providecommand{\newblock}{\relax}
\providecommand{\bibinfo}[2]{#2}
\providecommand{\BIBentrySTDinterwordspacing}{\spaceskip=0pt\relax}
\providecommand{\BIBentryALTinterwordstretchfactor}{4}
\providecommand{\BIBentryALTinterwordspacing}{\spaceskip=\fontdimen2\font plus
\BIBentryALTinterwordstretchfactor\fontdimen3\font minus
  \fontdimen4\font\relax}
\providecommand{\BIBforeignlanguage}[2]{{%
\expandafter\ifx\csname l@#1\endcsname\relax
\typeout{** WARNING: IEEEtran.bst: No hyphenation pattern has been}%
\typeout{** loaded for the language `#1'. Using the pattern for}%
\typeout{** the default language instead.}%
\else
\language=\csname l@#1\endcsname
\fi
#2}}
\providecommand{\BIBdecl}{\relax}
\BIBdecl

\bibitem{pantazis2009energy}
N.~A. Pantazis, D.~J. Vergados, D.~D. Vergados, and C.~Douligeris, ``Energy
  efficiency in wireless sensor networks using sleep mode tdma scheduling,''
  \emph{Ad Hoc Networks}, vol.~7, no.~2, pp. 322--343, 2009.

\bibitem{akyildiz2002survey}
I.~F. Akyildiz, W.~Su, Y.~Sankarasubramaniam, and E.~Cayirci, ``A survey on
  sensor networks,'' \emph{IEEE Communications magazine}, vol.~40, no.~8, pp.
  102--114, 2002.

\bibitem{al2015internet}
A.~Al-Fuqaha, M.~Guizani, M.~Mohammadi, M.~Aledhari, and M.~Ayyash, ``Internet
  of things: A survey on enabling technologies, protocols, and applications,''
  \emph{IEEE Communications Surveys \& Tutorials}, vol.~17, no.~4, pp.
  2347--2376, 2015.

\bibitem{p2}
G.~J. Pottie and W.~J. Kaiser, ``Wireless integrated network sensors,''
  \emph{Communications of the ACM}, vol.~43, no.~5, pp. 51--58, 2000.

\bibitem{vardhan2000wireless}
S.~Vardhan, M.~Wilczynski, G.~Portie, and W.~J. Kaiser, ``Wireless integrated
  network sensors (wins): distributed in situ sensing for mission and flight
  systems,'' in \emph{Aerospace Conference Proceedings, 2000 IEEE},
  vol.~7.\hskip 1em plus 0.5em minus 0.4em\relax IEEE, 2000, pp. 459--463.

\bibitem{akyildiz2002wireless}
I.~F. Akyildiz, W.~Su, Y.~Sankarasubramaniam, and E.~Cayirci, ``Wireless sensor
  networks: a survey,'' \emph{Computer networks}, vol.~38, no.~4, pp. 393--422,
  2002.

\bibitem{de2011energy}
G.~G. de~Oliveira~Brante, M.~T. Kakitani, and R.~D. Souza, ``Energy efficiency
  analysis of some cooperative and non-cooperative transmission schemes in
  wireless sensor networks,'' \emph{IEEE Transactions on Communications},
  vol.~59, no.~10, pp. 2671--2677, 2011.

\bibitem{rabaey2000picoradio}
J.~M. Rabaey, M.~J. Ammer, J.~L. Da~Silva, D.~Patel, and S.~Roundy, ``Picoradio
  supports ad hoc ultra-low power wireless networking,'' \emph{Computer},
  vol.~33, no.~7, pp. 42--48, 2000.

\bibitem{shih2001physical}
E.~Shih, S.-H. Cho, N.~Ickes, R.~Min, A.~Sinha, A.~Wang, and A.~Chandrakasan,
  ``Physical layer driven protocol and algorithm design for energy-efficient
  wireless sensor networks,'' in \emph{Proceedings of the 7th annual
  international conference on Mobile computing and networking}.\hskip 1em plus
  0.5em minus 0.4em\relax ACM, 2001, pp. 272--287.

\bibitem{hasan2011green}
Z.~Hasan, H.~Boostanimehr, and V.~K. Bhargava, ``Green cellular networks: A
  survey, some research issues and challenges,'' \emph{IEEE Communications
  surveys \& tutorials}, vol.~13, no.~4, pp. 524--540, 2011.

\bibitem{wang2010energy}
S.~Wang and J.~Nie, ``Energy efficiency optimization of cooperative
  communication in wireless sensor networks,'' \emph{EURASIP J. on Wireless
  Commun. and Network.}, vol. 2010, no.~1, p. 162326, 2010.

\bibitem{sadek2009energy}
A.~K. Sadek, W.~Yu, and K.~Liu, ``On the energy efficiency of cooperative
  communications in wireless sensor networks,'' \emph{ACM Transactions on
  Sensor Networks (TOSN)}, vol.~6, no.~1, p.~5, 2009.

\bibitem{alves2014outage}
H.~Alves, R.~D. Souza, and G.~Fraidenraich, ``Outage, throughput and energy
  efficiency analysis of some half and full duplex cooperative relaying
  schemes,'' \emph{Transactions on Emerging Telecommunications Technologies},
  vol.~25, no.~11, pp. 1114--1125, 2014.

\bibitem{nardelli2016maximizing}
P.~H. Nardelli, M.~de~Castro~Tom{\'e}, H.~Alves, C.~H. de~Lima, and
  M.~Latva-aho, ``Maximizing the link throughput between smart meters and
  aggregators as secondary users under power and outage constraints,'' \emph{Ad
  Hoc Networks}, vol.~41, pp. 57--68, 2016.

\bibitem{tome2016joint}
M.~C. Tom{\'e}, P.~H. Nardelli, H.~Alves, and M.~Latva-aho, ``Joint
  sampling-communication strategies for smart-meters to aggregator link as
  secondary users,'' in \emph{IEEE International Energy Conference (ENERGYCON),
  2016}.\hskip 1em plus 0.5em minus 0.4em\relax IEEE, 2016, pp. 1--6.

\bibitem{nardelli2012optimal}
P.~H. Nardelli, M.~Kaynia, P.~Cardieri, and M.~Latva-aho, ``Optimal
  transmission capacity of ad hoc networks with packet retransmissions,''
  \emph{IEEE Transactions on Wireless Communications}, vol.~11, no.~8, pp.
  2760--2766, 2012.

\bibitem{nardelli2014throughput}
P.~H. Nardelli, M.~Kountouris, P.~Cardieri, and M.~Latva-Aho, ``Throughput
  optimization in wireless networks under stability and packet loss
  constraints,'' \emph{IEEE Transactions on Mobile Computing}, vol.~13, no.~8,
  pp. 1883--1895, 2014.

\bibitem{wildman2014joint}
J.~Wildman, P.~H.~J. Nardelli, M.~Latva-aho, and S.~Weber, ``On the joint
  impact of beamwidth and orientation error on throughput in directional
  wireless poisson networks,'' \emph{IEEE Transactions on Wireless
  Communications}, vol.~13, no.~12, pp. 7072--7085, 2014.

\bibitem{haenggi2012stochastic}
M.~Haenggi, \emph{Stochastic geometry for wireless networks}.\hskip 1em plus
  0.5em minus 0.4em\relax Cambridge University Press, 2012.

\bibitem{weber2010overview}
S.~Weber, J.~G. Andrews, and N.~Jindal, ``An overview of the transmission
  capacity of wireless networks,'' \emph{IEEE Transactions on Communications},
  vol.~58, no.~12, pp. 3593--3604, 2010.

\bibitem{nardelli2015throughput}
P.~H. Nardelli, C.~H. de~Lima, H.~Alves, P.~Cardieri, and M.~Latva-aho,
  ``Throughput analysis of cognitive wireless networks with poisson distributed
  nodes based on location information,'' \emph{Ad Hoc Networks}, vol.~33, pp.
  1--15, 2015.

\bibitem{baccelli2011interference}
F.~Baccelli, A.~El~Gamal, and N.~David, ``Interference networks with
  point-to-point codes,'' \emph{IEEE Transactions on Information Theory},
  vol.~57, no.~5, pp. 2582--2596, 2011.

\bibitem{nardellimult2009i}
P.~H.~J. Nardelli, G.~T.~F. de~Abreu, and P.~Cardieri, ``Multi-hop aggregate
  information efficiency in wireless ad hoc networks,'' in \emph{IEEE Int.
  Conf. on Commun.}\hskip 1em plus 0.5em minus 0.4em\relax IEEE, 2009, pp.
  1--6.

\bibitem{Iran}
I.~Ramezanipour, P.~Nardelli, H.~Alves, and A.~Pouttu, ``Increasing the
  throughput of an unlicensed wireless network through retransmissions,'' in
  \emph{IEEE VTC-Spring}.\hskip 1em plus 0.5em minus 0.4em\relax
  ArXiv:1706.05813, 2018, pp. 1--5.

\end{thebibliography}


\end{document}